\RequirePackage[hyphens]{url}
\documentclass[aip, rsi, amsmath,amssymb, reprint, floatfix]{revtex4-1}

\usepackage{siunitx}
\usepackage{amsmath}
\usepackage{natbib}
\usepackage[breaklinks]{hyperref}
\usepackage{textcomp}
\usepackage[hyphens]{url}
\usepackage{breakurl}



\usepackage{mathtools}
\usepackage{graphicx}
\usepackage{dcolumn}
\usepackage{bm}
\usepackage{microtype}

\begin{document}


\title{Development of a SEMPA system for magnetic imaging with ns time resolution and phase-sensitive detection} 



\author{Daniel Sch\"onke}
\affiliation{Institut f\"ur Physik, Johannes Gutenberg Universit\"at Mainz, 55099 Mainz, Germany}

\author{Andreas Oelsner}
\affiliation{Surface Concept GmbH, 55099 Mainz, Germany}

\author{Pascal Krautscheid}
\affiliation{Institut f\"ur Physik, Johannes Gutenberg Universit\"at Mainz, 55099 Mainz, Germany}
\affiliation{MAINZ Graduate School, Johannes Gutenberg Universit\"at Mainz, 55099 Mainz, Germany}

\author{Robert M. Reeve}
\email[]{reeve@uni-mainz.de}
\affiliation{Institut f\"ur Physik, Johannes Gutenberg Universit\"at Mainz, 55099 Mainz, Germany}

\author{Mathias Kl\"aui}
\affiliation{Institut f\"ur Physik, Johannes Gutenberg Universit\"at Mainz, 55099 Mainz, Germany}
\affiliation{MAINZ Graduate School, Johannes Gutenberg Universit\"at Mainz, 55099 Mainz, Germany}


\date{\today}

\begin{abstract}
Scanning electron microscopy with polarization analysis is a powerful lab-based magnetic imaging technique offering parallel imaging of multiple magnetization components and a very high spatial resolution. However, one drawback of the technique is the long required acquisition time resulting from the low inherent efficiency of spin detection, which has limited the applicability of the technique to certain quasi-static measurement schemes and  materials with strong contrast. Here we demonstrate the ability to improve the signal-to-noise ratio for particular classes of measurement involving periodic excitation of the magnetic structure via the integration of a time-to-digital converter to the system and a digital lock-in detection scheme. The modified setup provides dynamic imaging capabilities using selected time windows and finally full time-resolved imaging with a demonstrated time resolution of better than 2 ns.
\end{abstract}


\maketitle 

\section{Introduction}

Magnetic imaging is a very important tool for the investigation of the underlying physical properties of magnetic systems and spintronic devices, for instance in order to correlate the electrical response of a system to the underlying static and dynamic spin configurations.  For magnetic devices the continued trend is towards smaller sizes and faster operation speeds, facilitated by improvements in lithographic and other fabrication techniques.  One currently active area of research towards these goals is the investigation of novel methods of switching nanoscale magnetic bits, for which imaging of the dynamic switching pathways is highly illuminative. Furthermore, new device architectures have been proposed based on the controlled propagation of magnetic domain walls\cite{Parkin190} or skyrmion quasi-particles \cite{fert2013skyrmions}, for which imaging of the detailed magnetic quasi-particle spin structure~\cite{krautscheid2016domain} and propagation is necessary. However, in order to adequately characterize these new systems it is necessary for magnetic imaging techniques to concurrently keep up with this trend the required high spatial and temporal resolution. 

Various magnetic imaging techniques exist with their various advantages, disadvantages and areas of applicability. Some of the most commonly used techniques with high spatial and temporal resolutions are provided at synchrotron X-ray facilities, with examples including scanning transmission x-ray microscopy \cite{kilcoyne2003interferometer} (STXM, $\sigma_{x}=20-30\,\mathrm{nm}$, $\sigma_{t}=50\,\mathrm{ps}$) and photoemission electron microscopy \cite{stohr1993element,guslienko2006magnetic} (PEEM, $\sigma_{x}=50\,\mathrm{nm}$, $\sigma_{t}=100\,\mathrm{ps}$). However for typical  studies the access to such large-scale facilities is severely limited and expensive and hence it is highly advantageous to complement such imaging with laboratory-based alternatives. Examples include Kerr microscopy \cite{fowler1954magnetic,McCord2015moke} based on the magneto-optical Kerr effect with a spatial resolution typically limited by the wavelength of the employed optical radiation ($\sigma_{x}=300\,\mathrm{nm}$, $\sigma_{t}=100\,\mathrm{fs}$) and Lorentz transmission electron microscopy \cite{chapman1999transmission,flannigan20124d} with a spatial resolution of $\sigma_{x}=10\,\mathrm{nm}$ and a possible temporal resolution down to $\sigma_{t}=1\,\mathrm{ps}$.

One other attractive lab-based technique is scanning electron microscopy with polarization analysis (SEMPA), developed by Koike et al in 1984 \cite{koike1984scanning}, which provides a high spatial resolution of below $30\,$nm~\cite{reeve2013magnetic} as standard and in favourable cases even down to a few nm~\cite{koike2013spin}. Additionally it offers a high-degree of surface sensitivity of around \SI{1}{nm}, in contrast to transmission measurements which probe the whole sample or optical techniques for which there is a much deeper penetration depth of the light. This is advantageous for the investigation of surface effects, but requires careful sample preparation and measurements need to be performed in ultrahigh vacuum. However, one main drawback of the technique to date has been the long required acquisition times. In general the scanning techniques like STXM and SEMPA are slower than the full-field techniques, but it is mainly the inherent low efficiency of the spin detector that is the reason for the long measurements in the case of SEMPA \cite{scheinfein1990scanning}. Furthermore this long acquisition time hampers dynamic imaging and renders conventional approaches, such as pump-probe imaging where the magnetic state is only captured at a single time delay for each excitation cycle, prohibitively time consuming. Only very recently has first progress towards time-resolved SEMPA been reported based on one particular system \cite{fromter2016time}, and hence widespread realization of dynamic SEMPA imaging is not yet available. Increasing the sensitivity of the detector or detection scheme is therefore highly desirable in order to detect materials with low contrast and save acquisition time. Here we report on our new implementation of an enhanced detection scheme for SEMPA based on the incorporation of a time-to-digital converter to the setup in order to record the arrival time of the individual electron scattering events from our spin-detector. We demonstrate that by employing a digital phase-sensitive detection (PSD) approach we are able to improve the signal to noise ratio for certain systems and measurements with periodically changing contrast and furthermore present and characterize some of the other measurement schemes afforded by the new system, including full dynamic imaging.

\section{Experimental Setup}
 
We use a commercial \textit{Omicron} system based on a Zeiss Gemini UHV Scanning Electron Microscope (SEM, similar to \cite{Jaksch1995}). It detects the spin polarization of secondary electrons which are emitted from a sample’s surface when hit by the unpolarized primary electron beam. Since this spin polarization is collinear with the magnetization of the probed surface region, local excitation of secondary electrons using a focused electron beam allows for the acquisition of a high-resolution image of the spin configuration by raster-scanning across the surface. The spin detection is based on spin polarized low energy electron diffraction (SPLEED) from a W crystal at an energy of \SI{102.5}{eV}~\cite{fromter2003miniaturized, fromter2011optimizing}. The electrons of the (2,0)-beams are then counted by two perpendicularly arranged channeltron electron-counter pairs. The calculated asymmetry is proportional to the polarization of the electron beam and thereby quantitatively related to the surface magnetization of the investigated sample~\cite{oepen2007scanning}. Using this detector geometry, the two in-plane spin polarization components can be measured at the same time by calculating the asymmetries of the counts:
\begin{align}
A_{1,2}=\frac{N_{1}-N_{2}}{N_{1}+N_{2}}.
\end{align}
The spin polarization is then given by the asymmetry divided by the sensitivity factor, which is analogous to the Sherman function describing the performance of Mott detectors. The counts are characterized by a Poisson distribution such that the statistical error of a polarization measurement with $N$ counts is given by
\begin{align}
\Delta P=\frac{1}{\sqrt{NS_{2}}}.
\end{align}
Meanwhile, the reflectivity, $R$, of the tungsten crystal is proportional to the total number of counts. Therefore the figure-of-merit $2RS^{2}$ can be used to quantify detector efficiency and to compare different spin detectors. Since typical values are about $10^{-4}$ it takes 10000 times longer to acquire a magnetic image with a comparable SNR to an intensity image for conventional SEM\cite{fromter2011optimizing,scheinfein1990scanning}. This low efficiency of the spin analyzer is the main challenge using SEMPA.
For magnetic imaging resolutions between $10-35\,\mathrm{nm}$ are routinely demonstrated and with a further optimized corrector in the electron optics a spatial resolution up to \SI{3}{nm} can be achieved via SEMPA\cite{koike2013spin}. To achieve the optimal resolution, however, noise sources from the environment have to be carefully controlled and eliminated. Electromagnetic noise can, for example, be caused by the \SI{50}{Hz} power supply system, its higher harmonics ($100, 150, 200\,\mathrm{Hz}$) and the radiofrequency coupling capacitively, inductively or resistively to the experiment\cite{lofink2014oberflachensensitive}. To reduce the influence of external fields our lab is fitted with an active compensation system (Integrated Dynamics Engineering EMI Compensation System MK 4) and mechanical vibrations are handled by a special damping system. To combat additional noise in the content of the present work a phase-sensitive detection system was implemented, which is applicable for measurements with periodical excitation of the magnetic state.

The main hardware addition to the system is a time-to-digital converter (TDC) which was customized for our particular requirements and integrated into our setup by Surface Concept (SC-TDC-1000/08 S). 

\begin{figure}
\includegraphics[width=0.9\columnwidth]{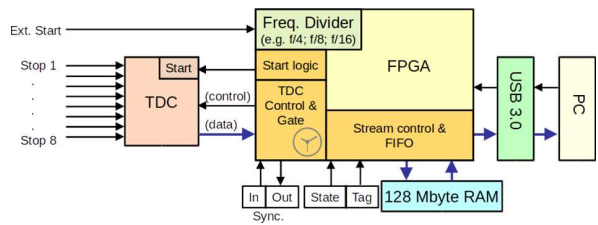}
\caption{\label{fig:TDCscheme} Schematic layout of the TDC. The detector output is fed into 4 ``Stop'' channels, the Ext. ``Start'' can be used as a reference as well as a fifth ``Stop'' input. The pixel trigger is provided to the ``Tag'' input.}
\end{figure}

The design of the SC-TDC-1000/08 S utilizes the high developed Time-to-Digital Converter (TDC) chip GPX of ACAM \cite{acamgpx} combined with a sophisticated control-, start-, and streaming-logic which is interfaced to any PC by a super-speed USB 3.0 connection. The device provides time measurements on 8 independent stop channels at a bin size of \SI{82.3}{ps} with a thermal stability of $<$ \SI{10}{fs/K}. The field programmable gate array (FPGA) enables convenient setups and a variable data stream handling from the TDC via USB. Fig.~\ref{fig:TDCscheme} shows the internal block scheme of the device including the various ``stop'', ``start'' and ``tag'' inputs. Parallel applied signals to different stop inputs can be detected without any time distance, however subsequent hits on any single stop input must have a delay of \SI{5.5}{ns} in order to be measured reliably. All single inputs provide a 32 fold multi-hit capability as an input buffer. The native GPX start input frequency is more restricted to below \SI{7}{MHz}, thus the logic provides a programmable frequency divider in order to operate on master frequencies up to \SI{100}{MHz}.  Although the high multi-hit capability of the GPX allows measurement of bursts of many pulses in a very short time interval (several tens of pulses within \SI{200}{ns}), the internal output registers and the maximum reading speed from the chip is limited, so that an average speed of up to 40 million results per second is possible uninterrupted and permanently. The TDC can be configured to measure as many hits as possible within these limits for either undefined or software specified acquisition intervals. This is required in the present application, since for typical materials and electron gun settings (\SI{3}{nA} beam current, \SI{3}{kV} acceleration voltage), channeltron count rates are up to \SI{4e6}{cts/s} per channel, so the device must be able to process an event rate of at least \SI{20}{MHz}. For the transfer to the computer USB 3.0 is used. The acquisition dwell times for time data from the TDC are settled within the FPGA by a quartz stabilized time gate in an interval from \SI{1}{ms} to \SI{1193}{h}. The start and end of measurement acquisition can be software triggered. For both modes, the start of the acquisition can also be hardware triggered by a TTL (transistor-transistor logic) ``Sync In'' signal and the end is signalized by a TTL pulse at the ``Sync. Out''. All stop times can be measured either in relation to any other previous stop result (including the same stop channel) or in relation to a periodical start at the terminal ``Ext. Start''. Stop time results can be synchronized to ``tag'' or ``state'' signals from the appropriate TTL inputs. The tag signal triggers an internal up-counter which starts with 0 at the start of each new acquisition block. The tag counter facilitates the pixel synchronization for the present application. The device can provide a start counter in parallel to each stop time result, as well as a tag result (either counter or timer or analog-digital-converter) and together with a 1 bit state result as a multiple coordinate system.
A complex FIFO (first-in-first-out buffer) design ensures a very low data loss probability of $<10^{-4}$ within the described limits. The user DLL (dynamic link library) controls the data handling and streaming for the user. 

\begin{figure}
\includegraphics[width=0.9\columnwidth]{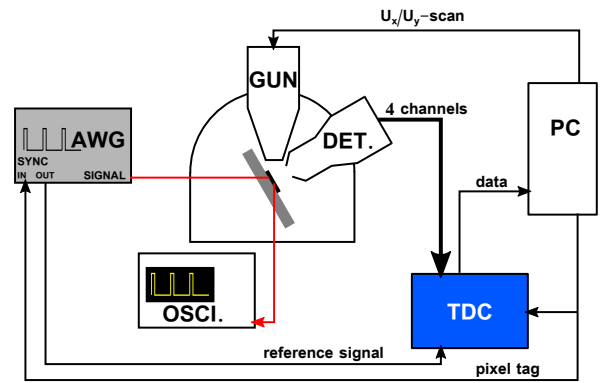}
\caption{\label{fig:SEMPAupgrade} Integrated setup of the new SEMPA system: The central part depicts the SEMPA investigation chamber including the electron gun and detector (det.). The scan is controlled from the PC by setting the electron beam deflection voltages ($U_x$, $U_y$). The measured counts are fed into the TDC and there they are assigned to the current pixel and time bin. The excitation of the device under investigation is performed using an arbitrary waveform generator (AWG), which is checked with an oscilloscope (osci.) and synchronized to the TDC.}
\end{figure}

The time resolution capability of the TDC (\SI{82.2}{ps}) is significantly better than required for our presently employed SPLEED detector system which is based on channeltron electron counters. These generate pulses of around $50\,$ns in length and $4\,$mV in height with risetimes of \SI{10}{ns} and can handle count rates of more than $10^7$ cps~\cite{manual}. The pulses are then passed through a discriminator and pre-amp arrangement in order to generate standard TTL compatible pulses of around $20\,$ns. Future upgrades to channelplates may be expected to provide improvement due to their shorter pulse spread, however, it has been demonstrated that electronic jitter due to the different flight times of the emitted electrons between the sample and the detector is likely to limit the ultimate achievable time resolution of the system to $\sim 600\,$ps for the present detector design and geometry~\cite{fromter2016time}. Nevertheless, larger scale design changes to the integrated system as a whole could provide further improvements \cite{fromter2016time}. The TDC detects the time of each individual electron pulse and depending on the programmed settings transfers selectively processed/packaged data to the computer for subsequent calculation and processing via software, enabling time-resolved measurements and the application of a digital phase-sensitive detection scheme. The required input connections to the TDC are the 4 channeltron low voltage TTL pulse lines. A pixel tag provided by the National Instruments (NI) PCIe 6343 card controlled via LabVIEW synchronizes with the scan frequency of the electron gun, i.e. the rate of the deflection voltage changes. For standard measurements the reference signal is also given by the pixel tag signal or in the case of measurements with periodic excitations by a synchronization signal provided by the arbitrary waveform generator (AWG). The deflection voltages for the electron beam U\textsubscript{x,y} for scanning the sample are also generated by the NI card and inputted to the electron gun and the change in U\textsubscript{x} serves as the master for pixel tag and – if wanted – excitation synchronization. We have implemented flexible scan-controls to allow for imaging particular regions of interest of arbitrary geometry within the field of view set by the microscope software. The UHV chamber sample stage has 4 electrical contacts to enable imaging while applying currents to, for example, a strip-line patterned on the sample for local magnetic field application. The excitation and reference signal for synchronization with the TDC imaging are provided by the AWG (Tektronix AWG4162) and monitored with an oscilloscope. A sketch of the integrated experimental setup is shown in Fig.~\ref{fig:SEMPAupgrade}. 

\subsection{General imaging mode}

The extended experimental setup can firstly be used for general magnetic imaging. For this, the pixel tag is fanned and also used as reference signal. In this mode the user sets a desired scan rate and the total counts within the corresponding dwell time are transferred from the TDC to the computer. Following each pixel, the beam deflection voltages are updated for the next position and the software generates a pixel tag to tell the TDC to start counting again. The next set of counts are then allocated to the next pixel within the image until the full scan is completed. Due to the long measurement time, drifts of the sample can play a significant role. To avoid consequent artifacts, many fast scan frames are acquired instead of only one long-time acquisition using an implemented multi-frame imaging mode. To prevent that the object of interest moves out of the field of view, a cross-correlation algorithm was included, which recalculates the region of interest after every frame using the sum of all 4 channeltron counts and updates the deflection voltages accordingly. This sum image shows the topography of the investigated area and has much better signal-to-noise ratio than the magnetic images and therefore is more suitable for pattern recognition. During post processing an additional cross-correlation matching is done to achieve best results. The measurement is controlled by a LabVIEW program communicating with the TDC via an integrated C++ software. To save more measurement time the user interface of the control allows to determine arbitrarily shaped regions of interest (ROIs), which might be useful for imaging structures such as magnetic rings where a large portion of the area is not of interest~\cite{krautscheid2016domain}.

\subsection{Time-resolved imaging}

The major advantage of the  new equipment is the possibilities it opens up for various types of new measurements that take advantage of the newly available time resolution. The first measurement scheme is a simple time resolved acquisition, where the excitation period is divided in a chosen number of time bins, $B$ \cite{fromter2016time}. For every scan over the sample, $B$ time bin images are acquired. In practice it is necessary to acquire over many pump-probe cycles in order to get sufficient SNR, however, unlike some pump-probe schemes we can simultaneously acquire the full time structure and so we acquire all time bin images for a given pixel within the one pump-probe cycle. We then repeat this for each pixel to generate all time bin images for each pixel. If the dwell time at one pixel is longer than a period of the excitation, the counts of the equivalent bins of every period are summed up. This means that neglecting the extra time required for a possible excitation pulse and any extra included duty cycle, the time for a movie with $B$ time bins is only increased by a factor of $B$  for a set SNR, unlike other acquisition schemes where you only acquire one time bin image per pump-probe cycle and the total time diverges as $B$ squared. The possible excitation-induced shifts (for instance due to Oersted fields generated by current pulses injected into the structures) within the time bin images of one single acquisition frame, in addition to thermal drifts can be corrected in a post processing step. For the drift correction in these modes the sum image of the first time window is used.

In the second mode you can selectively define only certain time windows in a period by choosing the offset from the reference pulse and the time window length allowing for e.g. measurement only during a current-pulse excitation of the sample. While this mode needs a longer measurement time to acquire images of comparable SNR depending on the fraction of ``acquisition on'' and ``acquisition off'' times, it significantly reduces the amount of data that is transferred from the TDC and ultimately saved and still captures the interesting time windows.

\subsection{Phase-sensitive measurement scheme}

\begin{figure}
\includegraphics[width=0.9\columnwidth]{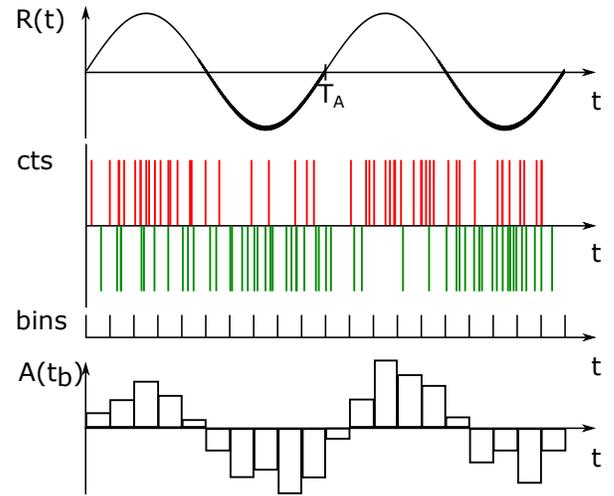}
\caption{\label{fig:LiAprinciple} Schematic working principle of the lock-in scheme: The sinusoidal excitation signal $R(t_{b})$ (top row) modulates the number of counts (cts) in the different channels as indicated in the second row where two opposing channeltron count trains are shown. The counts are collected into bins selected by the user, as shown in the 3rd row. Finally within a bin $b$ the asymmetry $A(t_{b})$ is calculated from the summed counts of this certain bin for each individual pixel.}
\end{figure}

\begin{figure}
\includegraphics[width=0.9\columnwidth]{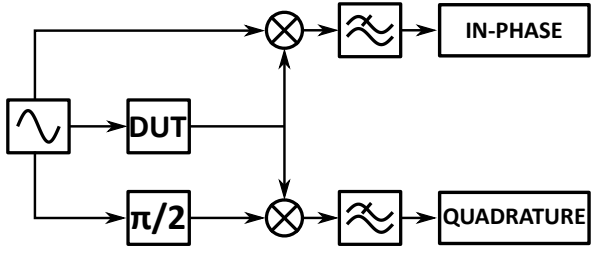}
\caption{\label{fig:LiA} General lock-in working principle: The sinusoidal excitation signal is applied to the device-under-test (DUT) and also used as a reference and mixed with the resulting signal of the experiment. In the dual-phase lock-in the reference is in addition shifted by 180 degrees and also mixed with the result. A low-pass filter is then applied to both branches. The amplitude is then calculated by the Pythagorean sum of the in-phase and quadrature component.}
\end{figure}

One completely new imaging mode that we implement in this work is a phase-sensitive signal detection \cite{braun2002computer} for SEMPA imaging. This mode is a further extension of the full period time resolved imaging and can be used in an extended mode by adding low-pass filtering also as a full lock-in detection(Fig.~\ref{fig:LiAprinciple}) as depicted in figure~\ref{fig:LiA}. This mode is in particular applicable if the magnetic signal is expected to be proportional to the excitation so that by taking the excitation frequency as a reference frequency, $\omega_{r}$, homodyne detection can be applied\cite{zhinst}. 

In this case the asymmetry oscillates with a period, $T_{A}$, which is equal to the period of the excitation which is split into a number of time bins, $B$, with bin length $\tau$, thus the $b$th bin is at a time $t_{b}=\tau(b-1)$ and $T_{A}=\tau B$. The resulting time-dependent asymmetry function $A(t_{b})$ is multiplied with a discretized complex reference signal $R(t_{b})\propto \mathrm{exp}(-i\omega_{r}t_{b})$. The two orthogonal contributions of this dual-phase down-mixing are the in-phase (IP) and quadrature (Q) components and are given in every pixel by 
 
\begin{align}
A_{\mathrm{IP}}(t_{b})=A(t_{b})\times\mathrm{sin}(\frac{b-0.5}{B}\times 2\pi)
\end{align} 
\begin{align}
A_{\mathrm{Q}}(t_{b})=A(t_{b})\times\mathrm{cos}(\frac{b-0.5}{B}\times 2\pi)
\end{align} 

In general the complex result of the dual-phase mixing is given by
\begin{align}
Z(t_{b})=A(t_{b})\times R(t_{b})
\end{align}
\begin{align}
Z(t_{b})=A(\mathrm{e}^{i(\omega_{A}-\omega_{r})t_{b}}+\mathrm{e}^{-i(\omega_{A}+\omega_{r})t_{b}})
\end{align}
If $\omega_{A}=\omega_{r}$, the first term becomes independent of time. The second term can be filtered out by a low-pass filter. For phase-sensitive detection this is done mathematically by simply averaging over the period. For the application of a specific low-pass filter implementation (digital RC, Butterworth, Chebychev,...) many periods of the asymmetry output have to be used for every pixel, since the filter needs a certain settling time. Afterwards the filtered data is integrated over time. In the end the complex result can be calculated by\cite{braun2002computer}
\begin{align}
Z=A_{\mathrm{IP}}+iA_{\mathrm{Q}}=\frac{2}{B}(\sum_{b=1}^{B}A_{\mathrm{IP}}(t_{b})+i\sum_{b=1}^{B}A_{\mathrm{Q}}(t_{b})).
\end{align}
Finally this result is transformed to the polar representation of $Z$ by the calculation of the absolute value 
\begin{align}
|Z|=A=\sqrt{A_{\mathrm{IP}}^{2}+A_{\mathrm{Q}}^{2}}
\end{align}
and the phase
\begin{align}
\theta=\mathrm{arctan}(\dfrac{A_{\mathrm{Q}}}{A_{\mathrm{IP}}}).
\end{align}
$\theta$ equals the phase shift of the signal compared to the excitation. This procedure is summarized in Figure~\ref{fig:LiA}.


\section{System Performance}

To test the system performance, various soft magnetic structures were patterned on a stripline (Cu(\SI{80}{nm})/Au(\SI{5}{nm}) or Cu(\SI{120}{nm})/Au(\SI{5}{nm})or Pt(\SI{20}{nm}) including squares, rectangles and disks with typical dimensions from $0.5-5\,$\textmu m. As ferromagnet, Ni\textsubscript{80}Fe\textsubscript{20} (Permalloy) with a thickness of \SI{10}{nm} or \SI{30}{nm} has been chosen. The samples were prepared by lift-off and then for imaging in-situ Ar milling was used to remove any surface contamination or surface oxide layer.

\subsection{Full period time-resolved imaging}

\begin{figure}
\includegraphics[width=0.9\columnwidth]{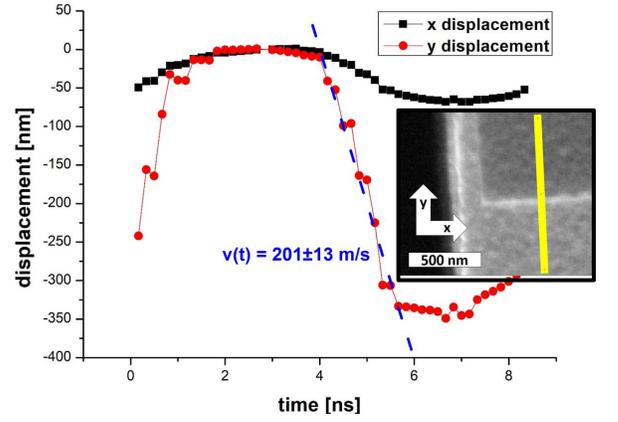}
\caption{\label{fig:tres} Determination of velocity of Py rectangle edge in $x$ and $y$ direction during excitation and determination of edge width (inset is showing the edge without excitation) for time resolution capability estimation.}
\end{figure}

\begin{figure}
\includegraphics[width=0.9\columnwidth]{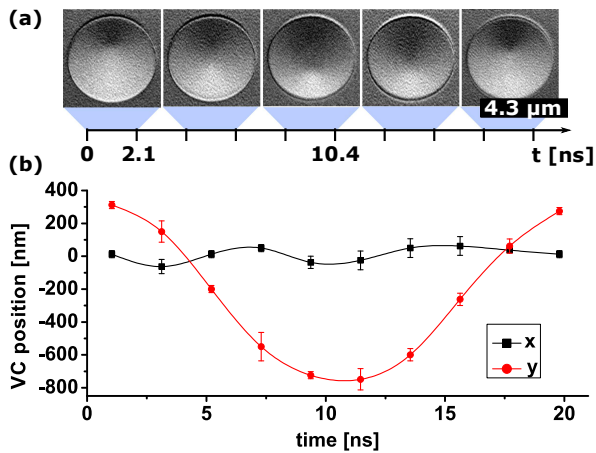}
\caption{\label{fig:trPy} (a) Vortex core (VC) gyration in a Py disk on Cu stripline due to oscillating current-induced Oersted field (48 MHz). A movie can be found here: http. (b) The trajectory shows the typical behaviour of an off-resonant excitation, with a large amplitude along the field axis and a smaller but detectable orthogonal oscillation. Clear changes on a few ns time scale are evident, confirming that magnetization dynamics imaging is possible using the previously estimated temporal resolution.}
\end{figure}

For the first two mentioned measurement modes, the time resolution $2\sigma_{t}$ is especially of interest. To determine this property we follow the protocol established by Fr\"omter et al.\cite{fromter2016time} is followed. At first, a static image of the edge of a $4$ by 3 \textmu m Py pad edge was acquired (inset of Fig.~\ref{fig:tres}) and the edge width $2\sigma_{0}$ measured by fitting a Gaussian profile to it. Then a second measurement using a sinusoidal excitation of 120 MHz was acquired with 10 bins per period. With a cross correlation algorithm, the positions of the edge were determined for every time frame in order to calculate velocity of the edge (Fig.~\ref{fig:tres}). The time frame image for the highest velocity $v(t)$ was taken to determine the edge width $2\sigma (t)$:
\begin{align}
2\sigma (t)=\sqrt{(2\sigma (t))^{2}+(2\sigma_{t}v(t))^{2}}
\end{align}
$2\sigma_{0}$ has been measured to be $45\pm30$ nm, $v(t)=200\pm13$ m/s and $2\sigma (t)=370\pm60$ nm, resulting in a time resolution of $1.8\pm 0.4$ nm.
The main limiting factor is expected to be the jitter in the channeltrons, which is on the order of nanoseconds and therefore higher than the calculated jitter in the travel time of the secondaries from the sample to the tungsten crystal\cite{fromter2016time}. Therefore the estimated time resolution could be further increased by exchanging the channeltrons by multi-channel plates with a lower velocity dispersion of the electrons \cite{fromter2016time}. 
As a demonstration of this performance, the field-driven vortex core motion of a permalloy (Py) disk of \SI{4.3}{\mu m} diameter and \SI{30}{nm} thickness was chosen. An oscillating Oersted field was generated by a sinusoidal current at 48 MHz flowing through the copper stripline (\SI{120}{nm} thickness capped with \SI{5}{nm} of Au) underneath the Py. Following Ampere`s law the Oersted field generated at the top of the stripline is estimated to be \SI{0.5}{mT}\cite{ntmdt}. The resonance frequency for this geometry is expected to be between 45 and 67 MHz from literature reports\cite{vansteenkiste2009stxmvc}. The period was oversampled by a factor of 10, resulting in a video of 10 time frames of \SI{2.1}{ns} each (5 frames shown in Fig.~\ref{fig:trPy}), multimedia file consisting of all frames: http). The total measurement time, including processing, was 211 minutes, of which around 50\%  was pure acquisition time. The computer processing time can be further reduced strongly by using the start input as reference instead of one of the 8 stop inputs, because the start signal is only handled by the hardware within the TDC.

\subsection{Time window imaging}
\begin{figure}
\includegraphics[width=0.9\columnwidth]{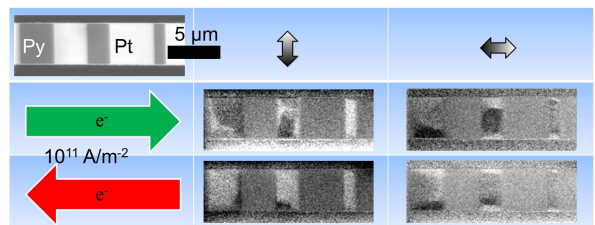}
\caption{\label{fig:gated} Gated imaging using two time windows for opposite current directions in Pt(\SI{20}{nm})/Py(\SI{10}{nm}) elements (\SI{50}{kHz} square wave, $\pm$\SI{0.1}{mA} amplitude). The change of the main magnetization direction is clearly visible for the two larger Py pads.}
\end{figure}

As a demonstration of the functionality of the second possible acquisition mode, Py rectangles of various sizes on a 20 nm Pt stripline were imaged during a rectangle pulse excitation of 50 kHz using 2 time intervals of interest, such that one is showing the magnetic structure during current flow to one direction while the other one belongs to the time during the reversed current direction. The absolute current amplitude is \SI{0.1}{mA} and the induced Oersted field is calculated to be \SI{1.25}{mT} at the center top of the Py structure.
Fig.~\ref{fig:gated} clearly shows how the magnetization direction is reversed by reversing the current through the wire in the two Py pads. Domains which are energetically favored i.e. aligned with the field increase their size at the expense of the others due to the influence of Zeeman energy. Only the narrowest structure remains largely unchanged for this value of the generated Oersted field due to the stronger shape anisotropy. Going to shorter time windows, one has to synchronize the whole setup very carefully to avoid runtime delays of voltage pulses and similar issues. For a full period acquisition this does not matter too much with exception of an observable phase shift, but in this mode it must be considered since under circumstances this phase shift can exceed the time window length. A possible way to determine the exact time of pulse arrival is to use a short acquisition using the first acquisition mode in order to check the shift of the observed object in the sum image followed by a recalculation of the time window of interest or the delay times of the system.

\subsection{Phase-sensitive detection}

\begin{figure}
\includegraphics[width=0.9\columnwidth]{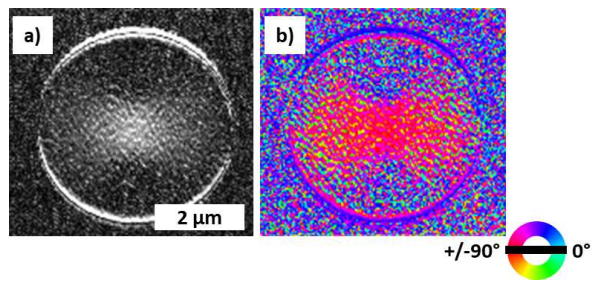}
\caption{\label{fig:LiAdisk} Amplitude (a) and color-coded phase image (b) of the time-resolved measurement (Fig.~\ref{fig:trPy}) revealing the excitation-caused magnetization changes in the Py disk.}
\end{figure}

As a first demonstration of the functionality of our phase-sensitive detection scheme we apply this scheme to the same vortex state excitation previously presented in Fig.~\ref{fig:trPy}. The results of the amplitude and phase are presented in Fig.~\ref{fig:LiAdisk}. In the amplitude image the strongest intensity corresponds to the region of the disk where the asymmetry signal is most correlated to the excitation frequency. In the phase image the pixel values are given by the phase shift between the signal and the excitation reference detected by the TDC. The less noisy an area of the image, the stronger the correlation of the signal to the excitation frequency. In Fig.~\ref{fig:LiAdisk} the region where the configuration changes has a high intensity in the amplitude image and is less speckled in the phase image. In regions where the magnetization does not change, the phase noise is indistinguishable from the region outside of the sample, as expected. Note that the phase can deviate from zero because of run-time delays of the pulses due to varying cable length. The white arcs in Fig.~\ref{fig:LiAdisk}a are artifacts corresponding to sample motion which has not been completely corrected for by the standard cross correlation algorithm.

\begin{figure}
\includegraphics[width=0.9\columnwidth]{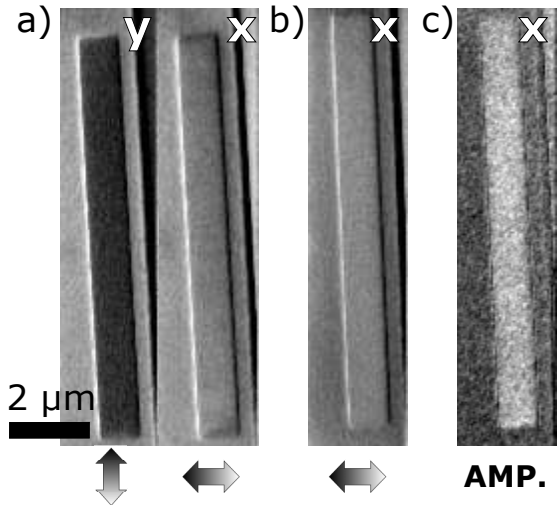}
\caption{\label{fig:LiA5} Comparison of the Py wire imaged (a) without current excitation (\SI{0}{mA} DC) , (b) with \SI{30}{mA} DC current excitation and standard static imaging and (c) with an AC excitation with a maximum current of \SI{30}{mA} and phase-sensitive detection (PSD, amplitude image). The current is flowing in the stripline parallel to the Py rectangle and the Oersted field points perpendicular to the long axis of the rectangle. (For the DC images the grey scale bar is the same for the $x$ component.)}
\end{figure}

The enhancement of the signal-to-noise that can be obtained by applying the lock-in detection to the acquired data was tested using a 1$\times$10 \textmu m$^{2}$ Py wire oriented along the long axis of the copper stripline. Due to the shape anisotropy, at remanence the magnetization in the wire is largely aligned along its long symmetry axis. Only at the top and bottom of the wire can domain patterns be observed due to flux-closure domains forming \cite{gomez1999domain}. Applying a current through the stripline results in an Oersted field that is directed perpendicular to this axis and which therefore competes with the shape anisotropy. The magnetization is therefore tilted by a certain angle, depending on the size of the Oersted field, which is in turn proportional to the current. As a result, for the x-component of magnetization no contrast is expected in the absence of current, however for finite current the contrast should increase in proportion to the amplitude and as such the lock-in imaging mode can be employed to detect this. 

To compare this technique to the usual imaging mode, two measurements with current were conducted: At first a DC current of \SI{30}{mA} was driven through the sample while imaging the magnetization tilt in the standard mode and secondly a sinusoidal current of the same amplitude was used and the phase-sensitive detection mode employed. The acquisition time for each of the measurements was \SI{20}{min}. For the lock-in measurement 32 bins per period were used. The dwell-time per pixel was \SI{0.1}{s} in both cases. The lock-in measurement was repeated with different excitation frequencies to investigate the frequency-dependent SNR. The SNR is found not to depend on the frequency used for the excitation/lock-in reference within the studied range of $10-1000\,\mathrm{kHz}$.
This detection scheme has to be interpreted very thoroughly because artificial signals can contribute to the resulting images. Two possible reasons for artifacts should be mentioned here: First, the drift correction is is unable to completely avoid image shifts, leading to a remaining excitation frequency coupled signal especially at the edges of structures as seen before for the discs (Fig.~\ref{fig:LiAdisk}). Another artificial contribution can occur due to the imperfect flatness of the asymmetry curve in dependence of the scattering potential of the SPLEED detector \cite{fromter2011optimizing}. For highly resistive devices the potential gradient over the sample can be significant, resulting in a position dependent scattering energy for the emitted electrons. In order to overcome this, it is necessary to either perform a complex correction of the data or alternatively measure using a virtual ground at the imaging position \cite{uhlig2009}. 


As can be seen by comparing the $x$ signals of Figs.~\ref{fig:LiA5}a-c, the lock-in scheme clearly enhances the signal to noise ratio in this case. Quantitatively we observe that an SNR enhancement of up to 5x can be achieved in the measured samples. It is, however, to be expected that the exact enhancement should depend very strongly on the specific measurement, sample and concretely on the exact frequency spectrum of the noise. For favorable cases, therefore, an even larger enhancement would be expected, however conversely we also expect and observe that for other measurements for instance where the noise occurs at the excitation frequency the SNR is not significantly different from the normal imaging mode based on equivalent acquisition settings.


\section{Conclusion}
In conclusion, we show that our upgraded system can, in addition to the standard SEMPA measurement mode, provide three further acquisition schemes: full-period time-resolved imaging, imaging during time windows and phase-sensitive detection. The achieved time-resolution is in the range of \SI{2}{ns}. For certain types of measurement, where the signal follows the excitation,an improvement in SNR of up to 5 has been observed using a phase-sensitive detection scheme. This work bodes well for future studies of materials with low contrast, which were previously inaccessible to SEMPA, as well as to new dynamic imaging modes. In particular, by combing the developments outlined here with recently improved efficient spin detectors and low jitter output electron detectors, we foresee that SEMPA could really compete with high-end synchrotron dynamic magnetic imaging techniques, yet in a convenient lab-based setting.

\section{Acknowledgements}
We thank Fabian Kloodt and Robert Fr\"omter for fruitful discussions. We acknowledge financial support of the SFB/TRR 173 Spin+X: spin in its collective environment (in particular Project B02) from the DFG, as well as the Graduate School of Excellence Materials Science in Mainz (GSC266).


%
%

%


\bibliography{SEMPA_tr_PSD}

\end{document}